\newif\ifeditorial 
\editorialfalse
\ifeditorial
\newcommand{\yaobin}[1]{\textcolor{red}{YS: #1}}
\else
\newcommand{\yaobin}[1]{}
\fi

\newif\ifsubmission   
\newif\iffinal       
\newif\iffull  
 
\submissionfalse
\finalfalse
\fulltrue
\documentclass[runningheads,a4paper]{llncs}
\usepackage{float} 
\usepackage{subcaption}

\usepackage{tablefootnote} % for table footnotes
\usepackage{verbatim} 
\usepackage{cmap} % makes ligature searchable & copy-able, to be used with [T1]{fontenc}
\usepackage[T1]{fontenc} % needs to come after \usepackage{cmap}
\usepackage{lmodern} % avoids bitmap fonts by [T1]{fontenc}
\usepackage{amssymb, amsmath}
\usepackage{graphicx, epsfig}
\usepackage[misc]{ifsym}
\usepackage{hyperref}
\usepackage{appendix}
\usepackage{tikz}
\usepackage{tkz-graph}
\usetikzlibrary{fit,arrows,calc,positioning}
\tikzset{stretch/.initial=1}
\usepackage{pgfplots}
\usepackage{pdfpages}
\usepackage{pgf}

\usepackage{enumerate}
\usepackage{attrib}
\usepackage{setspace}

\usepackage{algorithmicx}
\usepackage[noend]{algpseudocode}
\usepackage{algorithm}

\usepackage{framed, multicol}
\usepackage{ragged2e}

\usepackage{caption}
\usepackage{subcaption}
\usepackage{multirow}
\usepackage{supertabular}

\usepackage[T1]{fontenc}
\usepackage[utf8]{inputenc}
\usepackage{authblk}

\usepackage{xspace} %To be able to have clever spacings in commands
\usepackage{amssymb}
\usepackage{mathrsfs}

\algdef{SE}[DOWHILE]{Do}{doWhile}{\algorithmicdo}[1]{\algorithmicwhile\ #1}%
\algnewcommand\To{\textbf{to} }
\algrenewcommand\algorithmicprocedure{\textbf{Algorithm} }
\algrenewcommand\algorithmicfunction{\textbf{Game}}

\DeclareMathAlphabet{\mathsfsl}{T1}{cmss}{m}{sl}
\DeclareMathAlphabet{\mathsc}{OT1}{cmr}{m}{sc}
\DeclareMathAlphabet{\mathsl}{OT1}{cmr}{m}{sl}

\newcommand{\msb}[1]{\mathsf{msb}_{{#1}} }
\newcommand{\lsb}[1]{\mathsf{lsb}_{{#1}} }
\newcommand{\strtoint}{\mathsf{int}}
\newcommand{\inttostr}[1]{\mathsf{str}_{{#1}} }

\algrenewcommand\algorithmicprocedure{\textbf{Algorithm} }
\DeclareMathAlphabet{\mathsfsl}{T1}{cmss}{m}{sl}

\usepackage{array}
\usepackage{multirow}

\newcommand{\heading}[1]{\vspace{4pt}\noindent\textsc{#1}}

\newcommand{\bits}{\{0,1\}}

\newcommand{\getsb}{{\:{\leftarrow{\hspace*{-3pt}\raisebox{.75pt}{$\scriptscriptstyle n$}}}\:}}

\newcommand{\concat}{\:\|\:}

\iffinal

\else

\fi

\newcommand{\secref}[1]{Section~\ref{#1}\xspace}

\newlength{\saveparindent}
\newlength{\saveparskip}
\newcommand{\emptystring}{\{\}}

\newcommand{\GHASH}{\mathsf{GHASH}}

\newcommand{\abort}{\perp}

\newcommand{\len}[1]{|{#1}|}

\newcommand{\mode}[1]{{\sf{#1}}}

\newcommand{\gf}{\mathbf{GF}}

\begin{document}
\title{
	A Note on the Influence of a Zero Length Nonce on GCM and GMAC
}
%title remains to fill
\ifsubmission
\author{\ }
\institute{\ }
\else 
\author{Yaobin Shen
}
\institute{School of Informatics, Xiamen University, Xiamen, China\\
	\email{yaobin.shen@xmu.edu.cn} 
}
\fi

\maketitle
\ifsubmission
\vspace{-0.7in}
\fi

\begin{abstract}
	\iffalse
	HMAC and its variant NMAC are among the most widely used methods for keying a cryptographic hash function 
	to obtain a PRF or a MAC. 
	Yet, even after nearly three decades of research, 
	their generic PRF security still remains poorly understood, where the compression function of 
	the underlying hash function is treated as a black box and accessible to the adversary. 
	Although a series of works have exploited   
	compression function queries to mount generic attacks, 
	proving tight bounds on the generic PRF security of HMAC and NMAC 
	remains a challenging 
	open question until now.
%	by incorporating the influence of compression function queries 

	In this paper, we establish tight bounds on the generic PRF security of HMAC and NMAC. 
	Our bounds capture 
	the influence of the number of construction queries, the number of compression function queries, and 
	the maximal block length of a message on their security. The proofs are carried out in the 
	multi-user setting and the bounds hold regardless of the number of users. In addition, 
	we present matching attacks 
	to demonstrate that our bounds are essentially tight. Taken together, our results close a longstanding gap 
	in the generic PRF security analysis of HMAC and NMAC.
	\fi
	In this note, we show a simple attack that can recover the hash key of \mode{GCM} 
	and \mode{GMAC} by using a zero length nonce. After recovering the hash key, the 
	adversary can forge an arbitrary ciphertext or message as she wants. We note that 
	the ISO/IEC version of \mode{GCM} and \mode{GMAC} allows the nonce to be a zero length 
	string, while the NIST version of \mode{GCM} and \mode{GMAC} explicitly requires 
	the nonce to be at least one bit. Hence, our attack works for the ISO/IEC version and 
	cannot work for the NIST version. 
\end{abstract}

% !TEX root = ./main.tex

\section{Introduction}\label{sec:intro}

The Galois/Counter Mode (\mode{GCM}) and its specialization \mode{GMAC} are 
widely deployed authenticated encryption scheme and message authentication code scheme. 
They are designed by McGrew and Viega~\cite{DBLP:conf/indocrypt/McGrewV04}, 
and are based on the counter mode encryption and the polynomial hash function. 
They are specified in many international or national standards, 
including ISO/IEC~\cite{ISO-IEC-19772-gcm,ISO-IEC-9797-3-gmac}\yaobin{check the version of 3GPP later to see whether 3GPP allows an empty nonce. If no, then it is better no mention 3GPP}, TLS~\cite{tls1.2-dierks2008transport,tls1.3rescorla2018transport}, IPSec~\cite{viega2005use}, SSH~\cite{igoe2009aes}, WIFI (WPA3)~\cite{wifiwpa3}, NIST~\cite{nist-gcm/gmac-dworkin2007sp}, IEEE 802.1~\cite{ieee802.1}, and are widely used in practice. \yaobin{references for real-world applications, e.g., crypto libraries.}

The security of \mode{GCM} has been extensively evaluated from both the provable security side and the cryptanalysis side. 
McGrew and Viega~\cite{DBLP:conf/indocrypt/McGrewV04} proved that \mode{GCM} can achieve security up to $2^{n/2}$ data blocks, 
by assuming that 
the underlying $n$-bit block cipher is a secure PRP. Latex, Iwata et al.~\cite{DBLP:conf/crypto/IwataOM12} pointed out a flaw in the proof  
and fixed it. Their proof was then improved by Niwa et al.~\cite{niwa2015gcm} for a better security bound. \yaobin{should we mention the multi-user 
security result? It looks like not very necessary or relevant.} For the cryptanalysis aspect of \mode{GCM}, 
Ferguson~\cite{ferguson2005authentication} showed that forgery attacks and key recovery attacks become easier if the tag length is short. Joux~\cite{joux2006authentication} pointed out that 
a part of the secret key can be recovered if the nonce is reused. Handschuh and Preneel~\cite{handschuh2008key} systematically 
investigated weak keys of universal hash function, including \mode{GHASH}, a polynomial hash function employed 
in \mode{GCM}. The concept of weak keys for polynomial hash function was taken a step further by Saarinen~\cite{saarinen2012cycling} and 
he showed cycling forgery attacks against \mode{GCM} by exploiting weak keys. The cycling attacks was 
extended and formalized by Procter and Cid~\cite{procter2015weak} by introducing the notion of forgery polynomial. 
Abdelraheem et al.~\cite{abdelraheem2015twisted} showed a method to construct a forgery polynomial by using twisted polynomial from Ore rings. Note that none of these attacks violate the security claims for \mode{GCM} and \mode{GMAC}. 

In this note, we show a key recovery attack against \mode{GCM} and \mode{GMAC} by using single encryption query with a zero length nonce. 
This attack can recover the hash key of the polynomial hash function \mode{GHASH}. With the knowledge of this 
hash key, the adversary can arbitrarily forge a ciphertext against \mode{GCM} or arbitrarily forge a message 
against \mode{GMAC}. We note that the ISO/IEC version of \mode{GCM} and \mode{GMAC} allows the nonce to be 
a zero length string, while the NIST version requires the minimal length of a nonce to be 1 bit. Therefore, 
our attack can apply to the ISO/IEC version and cannot work for the NIST version. \yaobin{We also show the real-world applications of 
our attack...TBD}

\iffalse
\heading{Organization.} 
\secref{sec:prelim} introduces notation and definition. \secref{sec:hmac} studies the generic PRF security 
of \mode{HMAC} and shows the tightness of the proved bound. \secref{sec:nmac} studies the generic PRF security 
of \mode{NMAC} and discuss the tightness of the proved bound. 
%\secref{sec:concl} concludes the paper. 
\fi
% !TEX root = ./main.tex
\section{Description of GCM and GMAC}
\label{sec:gcm/gmac}
The Galois/Counter Mode (\mode{GCM}) is an authenticated encryption scheme that can be used to encrypt a plaintext 
and authenticate the resulting ciphertext together with associated data. The message authentication code scheme 
\mode{GMAC} can be regarded as a special 
case of \mode{GCM} where no plaintext is encrypted and only associated data is authenticated. 
Hence, in this section, we focus on the description of \mode{GCM}, as 
the specification of \mode{GMAC} can be obtained by putting the plaintext of \mode{GCM} as an empty string. 

We introduce some notations in order to describe \mode{GCM}. Let $\bits^*$ be the set of all finite bit strings, and 
$\bits^n$ the set of $n$-bit strings. Following the notation in~ISO/IEC~\cite{ISO-IEC-19772-gcm,ISO-IEC-9797-3-gmac}, 
Let the symbol $\emptystring$ be the string with zero length. The set of $n$-bit 
strings $\{0,1\}^n$ can also be regarded as $\gf(2^n)$, the finite field with $2^n$ elements. 
Let $\len{X}$ denote the length of a string $X$. $X0^*$ denotes the padding that appended with as few zero bits 
so that the length of string to be a multiple of $n$. 
%Let $\len{X}_n$ denote the $n$-bit encoding of the length of a string $X$. 
For a string $X$ and an integer $r$ with $\len{X}\geq r$, let $\msb{r}(X)$ be the string containing 
only the leftmost $r$ bits of string $X$, and $\lsb{r}(X)$ be the string containing only the rightmost $r$ bits of $X$. 
For non-negative integers $i$ and $r$ with $i \leq 2^{r}-1$, let 
$\inttostr{r}(i)$ be the $r$-bit binary representation of an integer $i$.
Let $\strtoint(X)$ be the non-negative integer representation of a string $X$.  
Concatenation of two strings $X$ and $Y$ is written as $X\concat Y$. For a string $X \in \bits^*$, 
let $X_1\concat X_2 \concat \ldots \concat X_{x-1}\concat X_x^* \getsb X$ that divides $X$ into $n$-bit blocks with the exception 
of the last block, where $\len{X_i} = n$ and $1\leq \len{X_x^*} \leq n$. 
Let $E:\bits^k \times \bits^n \to \bits^n$ be a block cipher. 
Let $E_K(X)$ be the block cipher encryption of the value $X \in \bits^n$ with the key $K \in \bits^k$. 
Let $X\cdot Y$ be the multiplication of two elements $X, Y \in \gf(2^n)$, $X \oplus Y$ 
the addition of $X$ and $Y$. 

The authenticated encryption and decryption algorithms of \mode{GCM} make use of a universal hash function called \mode{GHASH}. 
It takes as inputs a $n$-bit key, two arbitrary length strings, 
and gives a $n$-bit string as output. Given a $n$-bit key $H$, two arbitrary length (possibly empty) strings $A$ and $C$, 
where $A_1\concat A_2\concat \ldots \concat A_{a-1}\concat A_a^*$ $ \getsb A$ 
and $C_1\concat C_2\concat \ldots \concat C_{c-1}\concat C_c^* \getsb C$, 
$\mode{GHASH}(H,A,C)$ is the $n$-bit value $X_{a+c+1}$ that is defined as follows: 
\begin{enumerate}[(i)]
	\item $X_0 = 0^n$
	\item $X_i = (X_{i-1}\oplus A_i)\cdot H$, $1\leq i \leq a-1$
	\item $X_a = (X_{a-1}\oplus A_{a}0^*)\cdot H$
	\item $X_i = (X_{i-1}\oplus C_{i-a})\cdot H$, $a+1 \leq i \leq a+c-1$
	\item $X_{a+c} = (X_{a+c-1}\oplus C_{c}0^*)\cdot H$
	\item $X_{a+c+1} = (X_{a+c}\oplus \inttostr{n/2}(\len{A})\concat \inttostr{n/2}(\len{C}))\cdot H$ 
\end{enumerate}

We can now define the authenticated encryption algorithm of $\mode{GCM}$. It takes as inputs a secret key $K$, 
a variable length nonce $N$ which shall be distinct for every message, a plaintext $M$, and 
associated data $A$, and gives as outputs a ciphertext $C$ and a tag $T$. 
Here the length of a plaintext should be less than or equal to $2^{39} - 256$, and the length of 
associated data should be less than or equal to $2^{64}-1$.
The authenticated encryption 
algorithm is defined as follows: 
\begin{enumerate}[(i)]
	\item $H = E_K(0^n)$
	\item $Y_0 = N\concat 0^{31}\concat 1$ if $\len{N} = 96$; otherwise $Y_0 = \mode{GHASH}(H,\emptystring,N)$
	\item For $i = 1, 2,\ldots,m-1$, do the following two steps:
	\begin{enumerate}[(1)]
		\item $Y_i = \mathrm{inc}(Y_{i-1})$;
		\item $C_i = M_i \oplus E_K(Y_i)$.
	\end{enumerate}
	\item $Y_m = \mathrm{inc}(Y_{m-1})$
%	\item $C_m = M_m^* \oplus E_K(Y_m)[1:\len{M_m^*}]$
	\item $C_m^* = M_m^* \oplus \msb{\len{M_m^*}}(E_K(Y_m))$
	\item $T = \mode{GHASH}(H,A,C_1\concat C_2\concat \ldots \concat C_m) \oplus E_K(Y_0)$
\end{enumerate}
In the above definition, $\mathrm{inc}(X)$ treats the rightmost 32 bits of the string $X \in \bits^n$ as 
a non-negative integer, and increments this value modulo $2^{32}$. More informally, 
\[
\mathrm{inc}(X) = \msb{n-32}(X)\concat \inttostr{32}(\strtoint(\lsb{32}(X) + 1 \mod 2^{32})) \enspace. 
\]

Similarly, we can define the authenticated decryption algorithm of $\mode{GCM}$. It takes as inputs a secret key $K$, 
a variable length nonce $N$, a ciphertext $C$, associated data $A$, and gives as its output either the plaintext 
$M$, or the special symbol $\abort$ indicating that its inputs are not authentic. Formally, the authenticated 
decryption algorithm is defined as follows: 
\begin{enumerate}[(i)]
	\item $H = E_K(0^n)$
	\item $Y_0 = N\concat 0^{31}\concat 1$ if $\len{N} = 96$; otherwise $Y_0 = \mode{GHASH}(H,\emptystring,N)$
	\item $T' = \mode{GHASH}(H,A,C_1\concat C_2\concat \ldots \concat C_m) \oplus E_K(Y_0)$
	\item If $T' \neq T$, then return $\abort$
	\item For $i = 1, 2,\ldots,m-1$, do the following two steps:
	\begin{enumerate}[(1)]
		\item $Y_i = \mathrm{inc}(Y_{i-1})$;
		\item $M_i = C_i \oplus E_K(Y_i)$.
	\end{enumerate}
	\item $Y_m = \mathrm{inc}(Y_{m-1})$
	%	\item $C_m = M_m^* \oplus E_K(Y_m)[1:\len{M_m^*}]$
	\item $M_m^* = C_m^* \oplus \msb{\len{M_m^*}}(E_K(Y_m))$
\end{enumerate}

As mentioned above, the specification of \mode{GMAC} can be obtained by setting the plaintext and the ciphertext of \mode{GCM} 
as an empty string \emptystring.

The variable length nonce $N$ shall be unique for every message, otherwise \mode{GCM} and \mode{GMAC} will be vulnerable to the 
forgery attacks as described by Joux~\cite{joux2006authentication}. Note that in NIST version and the original version of GCM 
and GMAC~\cite{nist-gcm/gmac-dworkin2007sp,DBLP:conf/indocrypt/McGrewV04}, it explicitly requires 
that the length of a variable length nonce shall be between 1 and $2^{n/2} - 1$, 
while there is no such restriction in ISO/IEC version~\cite{ISO-IEC-19772-gcm,ISO-IEC-9797-3-gmac}. 
In ISO/IEC 9797-3~\cite{ISO-IEC-9797-3-gmac}, it allows the nonce to be of variable length without any requirement 
on the minimal length of the nonce. Hence, naturally it allows the nonce to be the zero length string $\emptystring$. 
In ISO/IEC 19772~\cite{ISO-IEC-19772-gcm}, there is a suggestion on the variable length of the nonce that should be 
between 1 and $2^{64}$. However, this suggestion appears in Appendix A.7 which is only informative and not a mandatory 
requirement. Although both the message and associate data can be an empty string, 
until now, we are not aware of any explanation why the nonce should be of length at least 1 bit or what 
will be the consequence if the nonce is a zero length string. 

\yaobin{Please check later whether we should mention that "Zero-length IVs nullify the authenticity guarantees of GCM" 
appears in Apple's CoreCrypto library and was discussed in crypto.stackexchange: https://crypto.stackexchange.com/questions/87886/what-are-the-implications-of-allowing-a-zero-length-gcm-nonce, although no one 
explain why they add this note. The answer in crypto.stackexchange somehow clarify that the zero-length IV should be fine 
via hashing in the length of the IV for variable IVs and for fixed-length IVs by setting the last bit to 1. However, 
the hashing of a zero length does not help to exclude this issue.}

% !TEX root = ./main.tex
\section{A Zero Length Nonce Attack against GCM and GMAC}
\label{sec:attacks}
\yaobin{There is a all zero nonce attack against GCM by Joux~\cite{joux2006authentication} to recover the hash key. Although 
the ideal is a bit similar, it is essential different. This all zero nonce attack cannot work for the current ISO/IEC version.}
\yaobin{Empty Nonce (A Zero Length Nonce, an empty nonce may be of length 1).}

In this section, we present an attack to recover the hash key of \mode{GCM} and \mode{GMAC}. This attack makes 
use of the zero length nonce\yaobin{It should be a zero length string instead of empty string, as an empty string may have a length of 1.} which is allowed in the ISO/IEC version of \mode{GCM}~\cite{ISO-IEC-19772-gcm} and \mode{GMAC}~\cite{ISO-IEC-9797-3-gmac}. 
This attack assumes that the adversary can observe the ciphertext corresponding 
to a zero length nonce (known ciphertext setting), or is granted the power to 
choose the nonce and set the nonce to be a zero length string (chosen plaintext setting). 

\heading{Zero length nonce attack.} In the ISO/IEC version of \mode{GCM} and \mode{GMAC}, 
they allow a variable length nonce 
and does not enforce the minimal length of a nonce. 
Hence, the adversary may observe the ciphertext with a zero length nonce, 
or is allowed to make a query with a zero length nonce. 
\yaobin{it may be more damaging as in the initial phase of some protocols, 
the sender may try to initialize the communication by sending a message 
with a zero length nonce to the receiver. But we may investigate further 
whether it occurs in which protocol.}
The attack works as follows. 
\begin{enumerate}
	\item The adversary observes the ciphertext and the tag $(C,T)$ corresponding to 
	an encryption query $(N,A,M)$ where $N$ is a zero length string $\emptystring$. 
	For the simplicity, assume that 
	both $A$ and $M$ are of one full-block length, but the attack works for $A$ and $M$ of 
	arbitrary length. Following the specification of \mode{GCM}, we have 
	\begin{enumerate}
		\item $Y_0 = \mode{GHASH}(H,\emptystring,\emptystring) = 0^n\cdot H = 0^n$
	%	$X_1 = A\cdot H$; $X_2 = (X_1\oplus C)\cdot H = A\cdot H^2 \oplus C\cdot H$; 
	%	$X_3 = (X_2 \oplus \inttostr{n/2}(n)\concat \inttostr{n/2}(n))\cdot H = A\cdot H^3 \oplus C\cdot H^2 
	%	\oplus (\inttostr{n/2}(n)\concat \inttostr{n/2}(n)) \cdot H$
		\item $T = \mode{GHASH}(H,A,C) \oplus E_K(0^n) = \mode{GHASH}(H, A, C) \oplus H = A\cdot H^3 \oplus C\cdot H^2 
		\oplus (\inttostr{n/2}(n)\concat \inttostr{n/2}(n)) \cdot H \oplus H$
	\end{enumerate}
	\item The above equation (b) is a cubic polynomial where the variable is $H$. By solving this equation, the adversary can obtain the hash key $H$. 
	\item After recovering the hash key $H$, the adversary can easily mount forgery attacks as follows. 
	\begin{itemize}
		\item By reusing the zero length nonce $N = \emptystring$, the adversary can then forge an arbitrary tuple of $(A', C', T')$ satisfying 
		\[
		T' = \GHASH(H, A', C') \oplus H \enspace.
		\]
		 $(N, A', C', T')$ is a valid tuple that can pass the authenticated decryption oracle.  
		\item The adversary can also forge by using any other observed tuple. 
		Once observed a tuple of $(N_1, A_1, C_1, T_1)$, she can create an new tuple of $(A', C', T')$ such that 
		\[
		\mode{GHASH}(H, A', C') \oplus T'= \mode{GHASH}(H, A_1, C_1) \oplus T_1 \enspace.
		\]
		The new tuple $(N_1,A',C',T')$ can pass the authenticated decryption oracle.
	\end{itemize} 
\end{enumerate}

It looks like the confidentiality may also be influenced to some extent. The adversary can collect the encryption of $Y_0'$ via $E_K(Y_0') = T' \oplus \mode{GHASH}(H,A',C')$ for each authenticated encryption query. 
In case that any one of counters used in the counter encryption mode collide with $Y_0'$ and 
denote by $C_i$ the corresponding ciphertext block, then the adversary can recover the 
plaintext via $M_i = C_i \oplus T' \oplus \mode{GHASH}(H,A',C')$. 

\heading{Remarks.} We have found that several real-world implementations accept zero-length nonces. To avoid exposing them to potential attacks, we do not identify them here and will notify the affected parties independently whenever possible.

\appendix

\bibliographystyle{abbrv}
\bibliography{refs}

@book{nist-gcm/gmac-dworkin2007sp,
  title={Sp 800-38d. recommendation for block cipher modes of operation: Galois/counter mode (gcm) and gmac},
  author={Dworkin, Morris J},
  year={2007},
  publisher={National Institute of Standards \& Technology}
}

@inproceedings{DBLP:conf/indocrypt/McGrewV04,
  author       = {David A. McGrew and
                  John Viega},
  title        = {The Security and Performance of the Galois/Counter Mode {(GCM)} of
                  Operation},
  booktitle    = {Progress in Cryptology - {INDOCRYPT} 2004, 5th International Conference
                  on Cryptology in India, Chennai, India, December 20-22, 2004, Proceedings},
  pages        = {343--355},
  year         = {2004},
  url          = {https://doi.org/10.1007/978-3-540-30556-9\_27},
  doi          = {10.1007/978-3-540-30556-9\_27},
  bibsource    = {dblp computer science bibliography, https://dblp.org}
}

@standard{ISO-IEC-19772-gcm,
  title        = {{Information Technology — Security Techniques — Authenticated Encryption}},
  organization = {International Organization for Standardization},
  institution  = {ISO/IEC},
  number       = {19772 : 2020},
  year         = {2020},
  note         = {},
  url          = {https://www.iso.org/standard/81550.html}
}

@standard{ISO-IEC-9797-3-gmac,
  title        = {{Information technology -- Security techniques -- Message Authentication Codes (MACs) -- Part 3: Mechanisms using a universal hash function}},
  organization = {International Organization for Standardization},
  institution  = {ISO/IEC},
  number       = {9797-3 : 2011},
  year         = {2011},
  note         = {},
  url          = {https://www.iso.org/standard/51619.html}
}

@article{joux2006authentication,
  title={Authentication failures in NIST version of GCM},
  author={Joux, Antoine},
  journal={NIST Comment},
  volume={3},
  year={2006}
}

@article{ferguson2005authentication,
  title={Authentication weaknesses in GCM},
  author={Ferguson, Niels},
  journal={NIST Comment},
  volume={},
  year={2005}
}

@techreport{tls1.2-dierks2008transport,
  title={The transport layer security (TLS) protocol version 1.2},
  author={Dierks, Tim and Rescorla, Eric},
  year={2008}
}

@techreport{tls1.3rescorla2018transport,
  title={The transport layer security (TLS) protocol version 1.3},
  author={Rescorla, Eric},
  year={2018}
}

@techreport{viega2005use,
  title={The use of Galois/counter mode (GCM) in IPsec encapsulating security payload (ESP)},
  author={Viega, John and McGrew, David},
  year={2005}
}

@techreport{igoe2009aes,
  title={AES Galois counter mode for the secure shell transport layer protocol},
  author={Igoe, Kevin and Solinas, Jerry},
  year={2009}
}

@standard{wifiwpa3,
  title        = {{WPA3 Specification Version 3.3}},
  organization = {Wi-Fi Alliance},
  year         = {2024},
  url          = {https://www.wi-fi.org/system/files/WPA3\%20Specification\%20v3.3.pdf}
}

@standard{ieee802.1,
  title        = {{IEEE Standard for Local and Metropolitan Area Networks Media Access Control
(MAC) Security}},
  organization = {IEEE Std 802.1AE-2006},
  year         = {2006},
  url          = {}
}

@inproceedings{DBLP:conf/crypto/IwataOM12,
  author       = {Tetsu Iwata and
                  Keisuke Ohashi and
                  Kazuhiko Minematsu},
  title        = {Breaking and Repairing {GCM} Security Proofs},
  booktitle    = {Advances in Cryptology - {CRYPTO} 2012 - 32nd Annual Cryptology Conference,
                  Santa Barbara, CA, USA, August 19-23, 2012. Proceedings},
  pages        = {31--49},
  year         = {2012},
  url          = {https://doi.org/10.1007/978-3-642-32009-5\_3},
  doi          = {10.1007/978-3-642-32009-5\_3},
  bibsource    = {dblp computer science bibliography, https://dblp.org}
}

@inproceedings{niwa2015gcm,
  title={GCM security bounds reconsidered},
  author={Niwa, Yuichi and Ohashi, Keisuke and Minematsu, Kazuhiko and Iwata, Tetsu},
  booktitle={International Workshop on Fast Software Encryption},
  pages={385--407},
  year={2015},
  organization={Springer}
}

@inproceedings{handschuh2008key,
  title={Key-recovery attacks on universal hash function based MAC algorithms},
  author={Handschuh, Helena and Preneel, Bart},
  booktitle={Annual International Cryptology Conference},
  pages={144--161},
  year={2008},
  organization={Springer}
}

@inproceedings{saarinen2012cycling,
  title={Cycling attacks on GCM, GHASH and other polynomial MACs and hashes},
  author={Saarinen, Markku-Juhani Olavi},
  booktitle={International Workshop on Fast Software Encryption},
  pages={216--225},
  year={2012},
  organization={Springer}
}

@article{procter2015weak,
  title={On weak keys and forgery attacks against polynomial-based MAC schemes},
  author={Procter, Gordon and Cid, Carlos},
  journal={Journal of Cryptology},
  volume={28},
  number={4},
  pages={769--795},
  year={2015},
  publisher={Springer}
}

@inproceedings{abdelraheem2015twisted,
  title={Twisted polynomials and forgery attacks on GCM},
  author={Abdelraheem, Mohamed Ahmed and Beelen, Peter and Bogdanov, Andrey and Tischhauser, Elmar},
  booktitle={Annual International Conference on the Theory and Applications of Cryptographic Techniques},
  pages={762--786},
  year={2015},
  organization={Springer}
}

\end{document}